# Thermal Emission Control via Twist Tuning of Embedded Eigenstates

Vladislav A. Chistyakov[1], Alex Krasnok[2*]

[1]*St Petersburg 197101, Russia*

[2]*Department of Electrical and Computer Engineering, Florida International University, Miami, FL 33174, USA*

*\*To whom correspondence should be addressed:* akrasnok@fiu.edu

**Abstract**

The field of thermal emission engineering shows great potential for various applications, such as lighting, energy harvesting, and imaging, using natural or artificial structures. However, existing structures face challenges in fabrication or do not provide the necessary degree of control over key parameters such as emission intensity, spectral composition, and angular distribution. To address these limitations, we propose a novel approach that leverages in-plane hyperbolic response, embedded eigenstates enabled by epsilon-near-zero, and exceptional tunability through twisting in α-MoO$_3$ heterostructures. By adjusting the twist angle, we can manipulate the system's properties, transforming it from a near-perfect reflector to a perfect absorber. This enables us to exert control over thermal emission power, spanning an order of magnitude. Furthermore, our research has uncovered a significant angular dependence of thermal emission, which varies with relative rotation.

*Introduction.* – Enhanced two-dimensional twisted bilayer materials showcase a vast array of striking electronic[1] and photonic[2–4] properties. By artfully adjusting the twist angle between two or more layers, exceptional control over electronic band structures is achieved. This has resulted in extraordinary phenomena such as magic-angle flat-band superconductivity[5,6], moiré exciton formation[7–12], and interlayer magnetism[13]. Recently, it was shown that combining analogous principles with the extreme anisotropy of natural or artificial hyperbolic materials enables control and manipulation of photonic dispersion in phonon or plasmon polaritons[2,14]. For instance, experiments unveiled tunable topological transitions from open (hyperbolic) to closed (elliptical) dispersion contours in α-phase molybdenum trioxide (α-MoO$_3$) bilayers[2]. These transitions occur when the rotation between layers attains a photonic magic twist angle. These groundbreaking findings broaden the horizons of twistronics and moiré physics, encompassing nanophotonics and polaritonics and paving the way for applications in nanoimaging, nanoscale light propagation, energy transfer, and quantum physics. While the focus of research was primarily on near-field effects, particularly surface polaritons in twisted heterostructures[3], there has been a significant shift in attention towards the far-field effects of moiré physics and complex heterostructures.

Thermal emission (TE) from heated objects is a significant far-field phenomenon that can be manipulated using natural or engineered structures, such as metamaterials, for applications in lighting, thermoregulation, energy harvesting, tagging, and imaging[15]. Controlling TE involves utilizing material and structural, optical resonances to enhance or suppress emission. Few-layer lossy thin films, for instance, enable control over absorption peaks' position and amplitude, and thus TE, potentially with a narrow bandwidth[16,17]. Photonic crystals offer substantial control over TE bandwidth, acting as passive filters reflecting TE at photonic-bandgap frequencies when integrated with emitting layers[18–20]. Metamaterials or metasurfaces, comprising arrays of resonant nanostructures with subwavelength separation, have been employed to achieve narrowband TE across visible, infrared, and terahertz ranges. Examples include cross-shaped metallic nanoantennas[21], metal stripes[22], slits in metallic surfaces[23], and semiconductor nanorods[24]. These structures facilitate the spectral-bandwidth engineering of TE. Directional control of TE adds a critical degree of freedom for various applications. Directional TE was initially observed in doped silicon gratings and later demonstrated with SiC [25], tungsten[26], and $SiO_2$[27] gratings in the infrared range. However, these structures present fabrication challenges or lack sufficient control over TE intensity, spectral composition, and angular distribution.

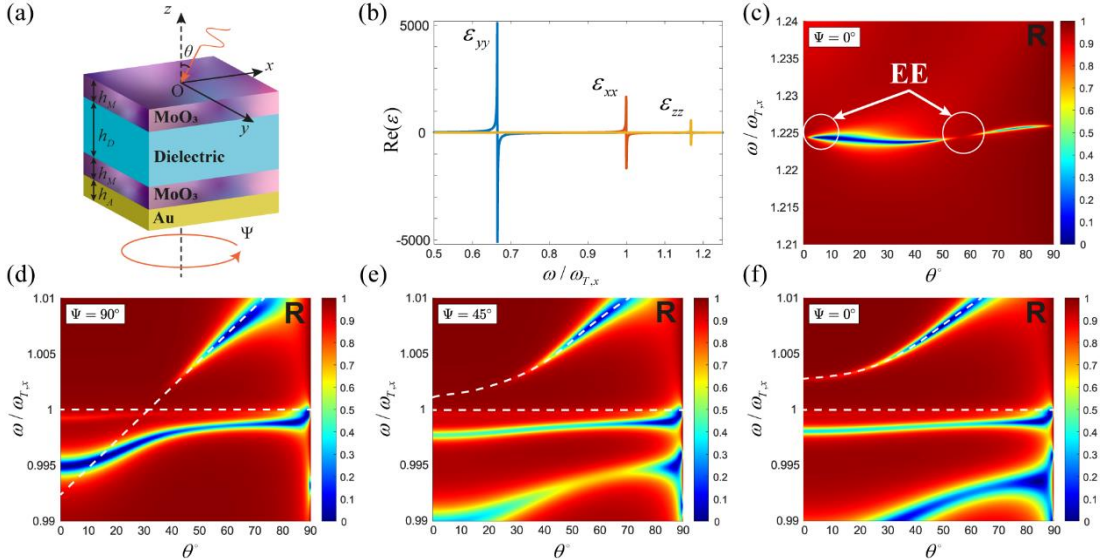

**Figure 1.** (a) Sketch of the structure comprising two monolayers of α-MoO₃ separated by Ge slab on Au substrate. (b) Permittivity of α-MoO₃ for low material losses $\Gamma'_n = \Gamma_n/10$. (c) Reflection spectrum around plasma frequency of $\varepsilon_{zz}$ as a function of normalized frequency and angle of incidence at a rotation angle $\Psi = 0°$. (d), (e), (f) Reflection spectra around resonant frequency $\omega_{T,x} = 820$ cm⁻¹ as a function of normalized frequency and angle of incidence at rotation angles $\Psi = 90°$, $\Psi = 45°$ and $\Psi = 0°$, respectively.

In this work, we propose a new method for thermal emission engineering at its extreme, which utilizes twisted hyperbolic heterostructures to make the approach more feasible. **Fig. 1a** provides a schematic representation of the model studied in this work. To demonstrate our approach, we analyze the commonly used α-MoO3 material, which possesses a natural hyperbolic resonance in the mid-infrared spectral range, eliminating the requirement for nanostructuring and lithography. α-MoO₃ monolayers can be conveniently fabricated at a scale of tens of micrometers. Axial heterostructure rotations are made easy by the weak van der Waals forces between layers, allowing for effortless tunability. Within the Reststrahlen band, ranging from 818 cm$^{-1}$ to 974 cm$^{-1}$, α-MoO$_3$ exhibits negative real permittivity along the y-direction while displaying positive real permittivity along the x-direction, **Fig.1b**. This striking anisotropy enables low-loss in-plane hyperbolic polariton propagation[2,28]. Notably, the hyperbolic response in α-MoO$_3$ occurs at a free-space wavelength of approximately 11 μm (frequency of ~900 cm$^{-1}$), corresponding to the black body thermal emission peak under normal conditions. Additionally, the optical response of α-MoO$_3$ bilayers within this frequency range presents an epsilon-near-zero (ENZ) regime, providing exceptional control over light scattering[29,30]. It has been shown that ENZ materials consisting of multiple layers can support embedded eigenstates (EE), also known as bound states in the continuum, with incredibly high Q-factors[31–34]. All these distinct properties make α-MoO$_3$ particularly compelling for thermal emission engineering applications. In this work, we synergistically combine the effects of in-plane hyperbolic response, ENZ-enabled embedded eigenstates with a high-Q response, and strong tunability facilitated by twisting in the feasible α-MoO$_3$ heterostructures to achieve unparalleled control over TE.

*Results and discussion.* – The structure depicted in **Fig.1a** was obtained through numerous optimization numerical simulations that we conducted. The structure comprises two optically thin layers of α-MoO$_3$ separated by a Ge slab on an Au substrate. The thicknesses of the top and bottom α-MoO$_3$ layers are $h_M = 270$ nm. $\alpha$-MoO3 is a hyperbolic material, and its permittivity tensor components are determined by the Lorentz equation[35]

$$\varepsilon_{nn} = \varepsilon_{\infty,n} \left( 1 + \frac{\omega_{L,n}^2 - \omega_{T,n}^2}{\omega_{T,n}^2 - \omega^2 - i\omega\Gamma_n'} \right), \tag{1}$$

where $\omega = 1/\lambda$ denotes the frequency in wavenumbers, and $n \in x, y, z$. The parameter values are[36] $\varepsilon_{\infty,x} = 4$, $\varepsilon_{\infty,y} = 5.2$, $\varepsilon_{\infty,z} = 2.4$, $\omega_{L,x} = 972$ cm$^{-1}$, $\omega_{L,y} = 851$ cm$^{-1}$, $\omega_{L,z} = 1004$ cm$^{-1}$, $\omega_{T,x} = 820$ cm$^{-1}$, $\omega_{T,y} = 545$ cm$^{-1}$, $\omega_{T,z} = 958$ cm$^{-1}$, $\Gamma_x = 4$ cm$^{-1}$, $\Gamma_y = 4$ cm$^{-1}$, and $\Gamma_z = 2$ cm$^{-1}$. Losses in the $\alpha$-MoO$_3$ material restrict the Q-factor of the EEs. In order to examine an almost ideal scenario and reduce the impact of losses, we will initially investigate material with lower levels of loss ($\Gamma_n' = \Gamma_n / 10$). The real parts of the permittivity tensor are depicted in **Fig.1b**. The case of real losses achievable in MoO$_3$ with the modern material qualities and fabrication techniques is discussed in what follows.

To confine the EE modes between the bilayered α-MoO$_3$, we separate them with a dielectric layer. We choose germanium (Ge) as the dielectric material with $\varepsilon_D = 16$, as it exhibits negligible losses in the considered frequency range. The dielectric thickness $h_D = 8.9$ μm is selected so that the Fabry-Pérot resonance of the layer corresponds to the frequency $\omega_{T,x}$, following the recent works[31–34]. The bottom layer is rotated relative to the top layer, with $\Psi = [0...90]°$ representing the angle of twist.

Additionally, the structure is situated on a thin gold (Au) substrate with a thickness of $h_A = 500$ nm. The results remain unaffected by the thickness of the Au substrate due to the nanometer-scale screening depth of Au within the frequency range of relevance.

To investigate the anisotropic structure, we employ the generalized T-matrix formalism method[36]. This algorithm can handle arbitrary anisotropic materials and does not produce discontinuous solutions. In order to demonstrate the existence of EEs in the system, we compute the full reflection coefficient $R^p = R^{pp} + R^{ps}$, where p- and s- denote polarization[37]. The structure is illuminated by a transverse-magnetic (TM or p)-polarized plane wave, with $\theta$ as the angle of incidence, **Fig.1a**. **Fig.1c** illustrates the reflection spectrum at the resonant frequency of the permittivity tensor component $\varepsilon_{zz}$, with a fixed twist angle of $\Psi = 0°$. The EEs states manifest themselves by the vanishing resonance line in the reflection spectrum at a normal angle of incidence $\theta = 0°$ and $\theta = 59°$. The reflection coefficient approaches unity at these EE states, and tunneling ceases. Losses in the material render these states the quasi-EEs[29].

A particularly intriguing scenario arises when one EEs is tuned upon layers rotation. To examine this case, we analyze the reflection spectrum around resonant frequency $\omega_{T,x} = 820$ cm$^{-1}$ as a function of normalized frequency and angle of incidence at different rotation angles. We consider the reflection spectrum at the twist angle of $\Psi = 90°$, **Fig.1d**. The spectrum shows the weak coupling regime, where the mode branches of α-MoO3 material resonance (horizontal branch) and the photonic mode of Ge slab (inclined branch) are visible[29,38,39]. By decreasing the twist angle, the system transitions into a strong coupling mode, as shown in **Fig.1e**. Hybrid eigenstates of the interacting subsystems arise in this scenario, manifesting Rabi splitting in the spectrum[40]. When the twist angle reaches $\Psi = 0°$, the mode splitting attains its maximum value, as seen in **Fig.1f**. These results unveil the potential to control the interaction energy between the subsystems by manipulating the twist angle between the layers.

We examine the reflection spectrum's poles and zeros in the complex frequency domain to delve deeper into the interaction between system modes and the process by which the system transits to the EE state. **Fig. 2a** illustrates the reflection spectrum as a function of imaginary normalized frequency and real frequency for an incidence angle of $\theta = 45°$ and a twist angle of $\Psi = 0°$. This condition corresponds to the upper branch mode in **Fig.1f**. Due to the finite Q-factor of the mode due to radiation and dissipation, the pole and zero associated with the mode are observed in the lower complex frequency plane. The limiting case of a Hermitian lossless system would be the EE, where the zero and the pole coalesce on the real axis, resulting in their coherent cancellation[29,33,41,42]. However, this does not occur due to the system's non-Hermitian nature. **Fig.2b** shows that the pole and zero possess opposite topological charges, suggesting they could merge if material losses were ignored[34]. **Fig. 2c** presents the trajectories of the singularities as the incidence angle $\theta$ increases from 0° to 90° while maintaining a fixed twist angle of $\Psi = 0°$. At the normal incidence angle, the pole and zero coalesce, forming a quasi-EE state, in agreement with the results in **Fig.1f**. As the incidence angle increases, the pole and zero diverge, causing the singular point to vanishing. **Fig. 2d** depicts the trajectories of the poles and zeros for a fixed angle of incidence $\theta = 45°$ as the twist angle $\Psi$ changes from $0°$ to $90°$. The pole and zero move downward along the imaginary and real frequency axes as the twist angle increases, indicating a reduction in the coupling energy and an increase in dissipative losses within the system, in agreement with the results in **Fig.1d**.

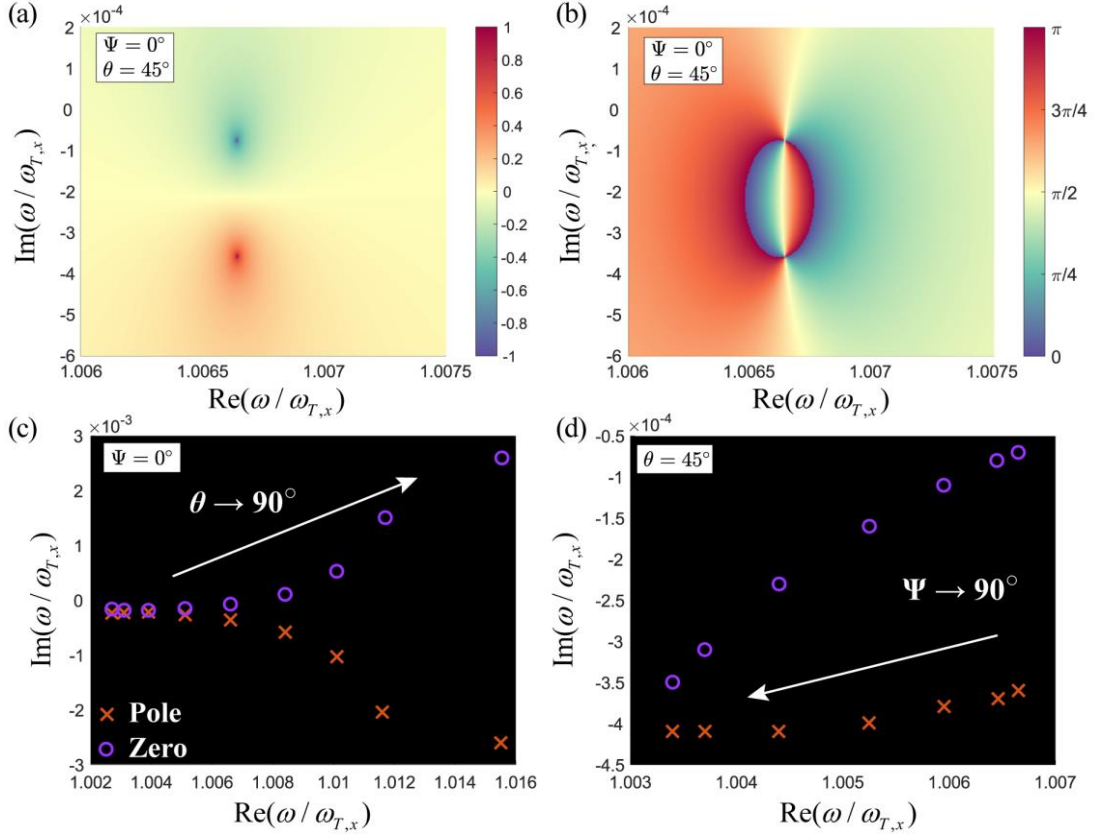

**Figure 2.** (a),(b) Reflection amplitude (a) and phase (b) in the complex frequency plane at an incidence angle of $\theta = 45°$ and a twist angle of $\Psi = 0°$ demonstrating the zero (blue) and pole (red). (c), (d) Positions of zeros and poles of the reflection in the complex frequency plane at fixed $\Psi = 0°$ ($\theta = 45°$) and varying $\theta$ ($\Psi$).

Until now, the results have only been presented for low-dose scenarios. However, it's important to recognize that higher losses can significantly impact the proposed system. As the losses increase, the Q-factor of the quasi-EE resonances decreases, which causes a downward shift in the zero and pole modes in the complex plane. In this case, the suggested structure can function as an ultranarrow perfect absorber with controlled thermal emission enhancement via the $MoO_3$ layer rotations.

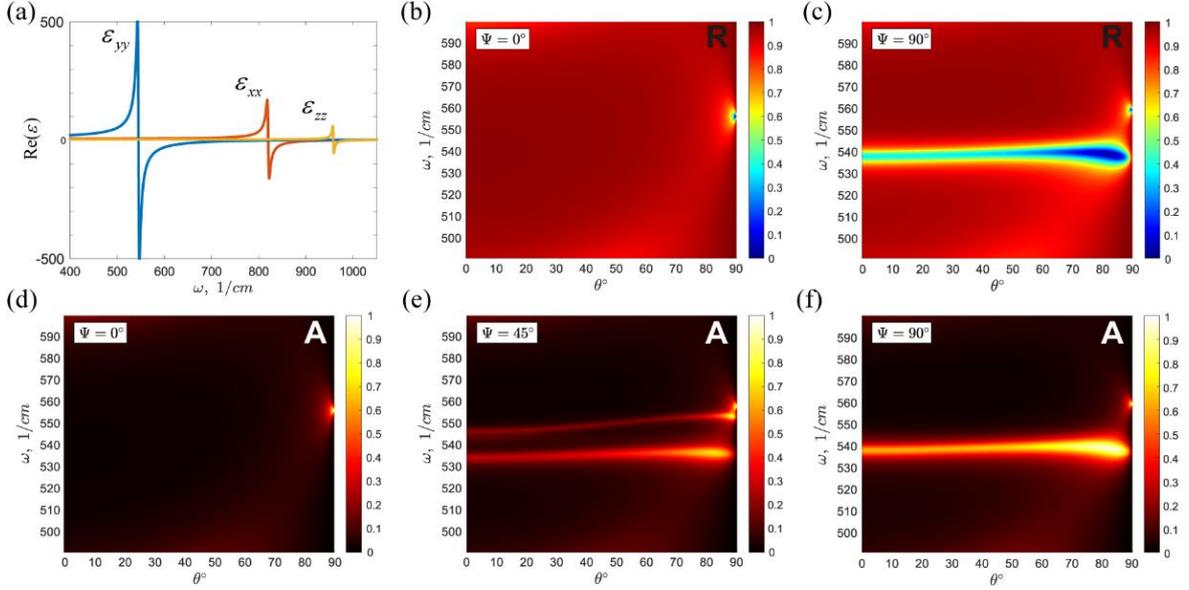

**Figure 3.** (a) Permittivity of α-MoO₃ with realistic losses. (b), (c) Reflection spectrum as a function of frequency around $\omega_{T,y}$ and angle of incidence at the rotation angle of $\Psi = 0°$ and $\Psi = 90°$, respectively. (d)-(e) Absorption spectrum as a function of frequency $\omega$ around $\omega_{T,y}$ and angle of incidence $\theta$ at the rotation angles of $\Psi = 0°$, $\Psi = 45°$ and $\Psi = 90°$, respectively.

**Fig.3a** illustrates the real part of the permittivity tensor components for realistic dissipative losses, denoted by $\Gamma'_n = \Gamma$ [35]. The parameters for the scheme proposed in **Fig.1a** remain the same. We investigate the tuning phenomenon by selecting the frequency range around the material resonance $\omega_{T,y}$. **Fig.3b** presents the reflection spectrum as a function of frequency ω and angle of incidence θ at a rotation angle $\Psi = 0°$. Exciting the structure with a TM-polarized plane wave ensures that the $\varepsilon_{yy}$ component mode remains unexcited. As the twist angle increases, this mode of α-MoO₃ gets excited, significantly affecting the system's response. Minimal reflection (maximal absorption) is achieved at $\Psi = 90°$ and an angle of incidence $\theta = 80°$, see **Fig.3c**.

To investigate the impact of absorption, we analyze how the absorption spectrum of the structure changes as the twist angle varies. The absorptivity of the complete multilayer is defined as $A^p = 1 - R^p - T^p$ [37]. When the relative twisting angle of the layers is $\Psi = 0°$, the polariton mode of α-MoO₃ remains closed, yielding minimal absorption across the entire frequency range, **Fig.3d**. **Fig.3e** vividly demonstrates the interaction between the phonon-polariton mode of α-MoO₃ and the Fabry-Perot mode of the dielectric at a relative rotation angle of $\Psi = 45°$. The interaction led to the formation of two distinct modes visible in the absorption spectrum. Finally, **Fig.3f** reveals near-perfect absorption in proximity to the α-MoO₃ mode at a twisting angle of $\Psi = 90°$. This interaction exhibits narrowband spectral and spatial properties, highlighting the significance of the ENZ-resonator-ENZ configuration in controlling the absorption magnitude through the relative rotation of α-MoO₃ layers.

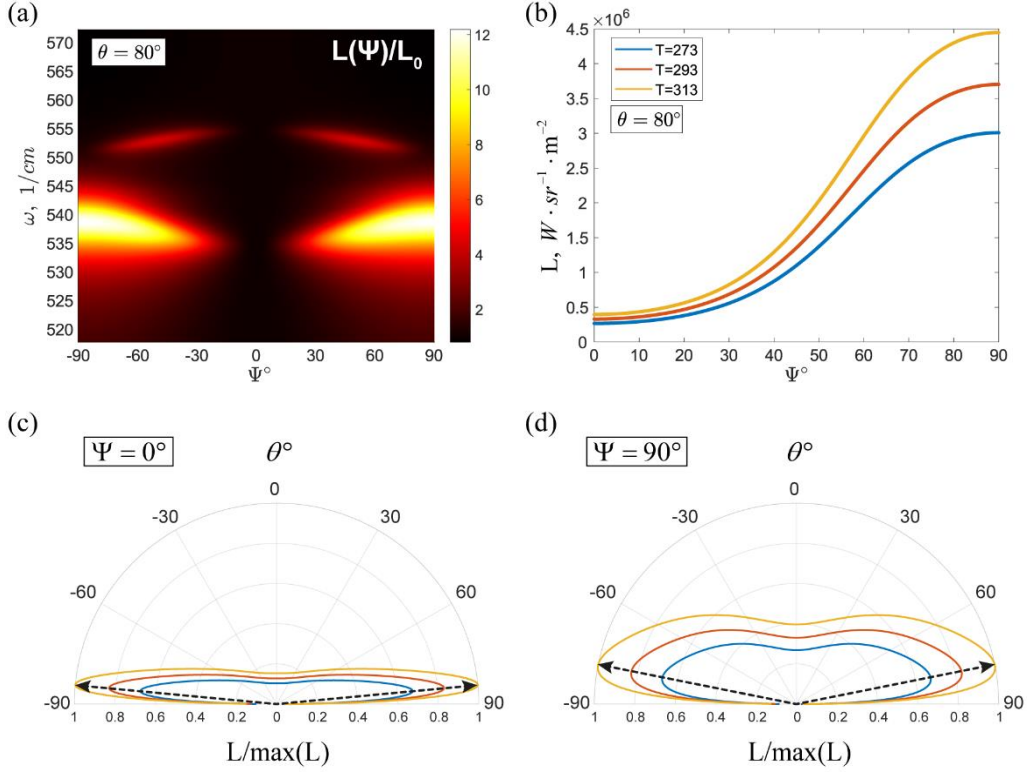

**Figure 4.** (a) Tunability spectrum $L(\omega,\Psi)/L_0(\omega,\Psi)$ versus frequency and rotation angle at $T=293$ K and $\theta=80°$. (b) Spectral radiance as a function of the twist angle at the angle of incident $\theta=80°$ and $\omega=540$ cm$^{-1}$ for temperatures $T=273$, $293$, and $313$ K. (c),(d) Angular distribution of spectral radiance at the twist angles of $\Psi=0°$ (c) and $\Psi=90°$ (d), and $\omega=540$ cm$^{-1}$ for temperatures $T=273$, $293$, and $313$ K.

Due to its ability to reconfigure the absorption coefficient, exceptional spectral selectivity, and operation frequency within the infrared range (17-20 μm), the proposed structure is an ideal candidate for developing a reconfigurable far-field thermal emitter[15]. To showcase the capabilities of this approach, it is crucial to analyze the spectral radiance characteristic $L(\theta,\omega,\Psi,T)$, defined as the radiance of a surface per unit frequency. In the far-field regime and at a fixed temperature, we can use the concept of emissivity $E(\theta,\omega,\Psi,T)$, to define the spectral radiance as $L(\theta,\omega,\Psi,T)=E(\theta,\omega,\Psi,T)\cdot L_{BB}(\omega,T)$. Here $L_{BB}(\omega,T)$ is Planck's law for the spectral radiance of a black body:

$$L_{BB}(\omega,T)=\frac{\hbar\omega^3}{4\pi^3 c^2}\frac{1}{e^{\hbar\omega/k_B T}-1}, \qquad (2)$$

where $c$ is the speed of light, $\hbar$ is Planck's constant, and $k_B$ is the Boltzmann constant. For an arbitrary body in thermodynamic equilibrium, emissivity is equal to absorptivity, as stated by Kirchhoff's law, $E(\theta,\omega,\Psi,T)=A(\theta,\omega,\Psi,T)$. This results in the following working formula for the spectral radiance of our system:

$$L(\theta,\omega,\Psi,T) = \frac{\hbar\omega^3}{4\pi^3 c^2} \frac{1}{e^{\hbar\omega/k_B T}-1} A(\theta,\omega,\Psi,T) \,. \tag{3}$$

The tunability of a multilayer structure refers to the spectral radiance ratio between a changed twist angle and a fixed angle $\Psi=0$, $L(\omega,\Psi)/L_0(\omega,\Psi)$. The results for the tunability spectrum $L(\omega,\Psi)/L_0(\omega,\Psi)$ versus frequency and rotation angle at $T=293$ K and $\theta=80°$ are presented in **Fig.4a**. Remarkably, the tunability coefficient reaches a value of 12 at $\Psi=\pm 90°$, $\theta=80°$ and $\omega=540$ cm$^{-1}$, which corresponds to the resonant frequency for $\varepsilon_{yy}$. This tunability enables the modification of the system's characteristics from a perfect reflector to a perfect absorber by adjusting the twist angle. **Fig.4b** presents the spectral radiance as a function of the twist angle at the angle of incident $\theta=80°$ and $\omega=540$ cm$^{-1}$ for temperatures $T=273$, $293$, and $313$ K. Variation in the relative angle results in a tenfold change in spectral radiance. This indicates that the thermal emitter has two distinct operating modes: intense emission and negligible emission. The angular distributions of spectral radiance at the twist angles of $\Psi=0°$ and $\Psi=90°$ for $\omega=540$ cm$^{-1}$ and temperatures $T=273$, $293$, and $313$ K are presented in **Fig.4c,d**, respectively. We not only notice a significant angular dependence of the TE, which increases in magnitude as the temperature rises but also a noteworthy alteration in the primary lobe's direction of this radiation upon relative rotation. This allows for manipulating a specific distribution of TE by twisting the structure.

*Conclusions.* – In conclusion, this study presents a novel approach to thermal emission engineering using MoO$_3$ heterostructures, overcoming limitations of existing structures by skillfully combining in-plane hyperbolic response, epsilon-near-zero (ENZ) enabled embedded eigenstates, and exceptional tunability through twisting. By adjusting the twist angle, we can transform the system from an ideal reflector to a perfect absorber, achieving an order of magnitude control over thermal emission amplitude. This innovative method of manipulating thermal radiation holds immense potential for various energy conversion applications. Our findings demonstrate that the tunability of a multilayer structure can be efficiently controlled via twist angle adjustment, creating a versatile system with promising implications for future research and development in energy conversion fields.

**Acknowledgments**

The authors thank Prof. Andrea Alú for fruitful discussions. The majority of this work was conducted prior to February 24th, 2022.

**References**

(1) Castellanos-Gomez, A.; Duan, X.; Fei, Z.; Gutierrez, H. R.; Huang, Y.; Huang, X.; Quereda, J.; Qian, Q.; Sutter, E.; Sutter, P. Van Der Waals Heterostructures. *Nat. Rev. Methods Prim.* **2022**, *2* (1), 58. https://doi.org/10.1038/s43586-022-00139-1.

(2) Hu, G.; Ou, Q.; Si, G.; Wu, Y.; Wu, J.; Dai, Z.; Krasnok, A.; Mazor, Y.; Zhang, Q.; Bao, Q.; Qiu, C.-W.; Alù, A. Topological Polaritons and Photonic Magic Angles in Twisted α-MoO3 Bilayers. *Nature* **2020**, *582* (7811), 209–213. https://doi.org/10.1038/s41586-020-2359-9.

(3) Zhang, Q.; Hu, G.; Ma, W.; Li, P.; Krasnok, A.; Hillenbrand, R.; Alù, A.; Qiu, C.-W. Interface Nano-Optics with van Der Waals Polaritons. *Nature* **2021**, *597* (7875), 187–195. https://doi.org/10.1038/s41586-021-03581-5.

(4) Jiang, Y.; Chen, S.; Zheng, W.; Zheng, B.; Pan, A. Interlayer Exciton Formation, Relaxation, and Transport in TMD van Der Waals Heterostructures. *Light Sci. Appl.* **2021**, *10* (1), 72. https://doi.org/10.1038/s41377-021-00500-1.

(5) Cao, Y.; Fatemi, V.; Fang, S.; Watanabe, K.; Taniguchi, T.; Kaxiras, E.; Jarillo-Herrero, P. Unconventional Superconductivity in Magic-Angle Graphene Superlattices. *Nature* **2018**, *556* (7699), 43–50. https://doi.org/10.1038/nature26160.

(6) Cao, Y.; Chowdhury, D.; Rodan-Legrain, D.; Rubies-Bigorda, O.; Watanabe, K.; Taniguchi, T.; Senthil, T.; Jarillo-Herrero, P. Strange Metal in Magic-Angle Graphene with near Planckian Dissipation. *Phys. Rev. Lett.* **2020**, *124* (7), 076801. https://doi.org/10.1103/PhysRevLett.124.076801.

(7) Wu, F.; Lovorn, T.; MacDonald, A. H. Topological Exciton Bands in Moiré Heterojunctions. *Phys. Rev. Lett.* **2017**, *118* (14), 147401. https://doi.org/10.1103/PhysRevLett.118.147401.

(8) Yu, H.; Liu, G.-B.; Tang, J.; Xu, X.; Yao, W. Moiré Excitons: From Programmable Quantum Emitter Arrays to Spin-Orbit–Coupled Artificial Lattices. *Sci. Adv.* **2017**, *3* (11). https://doi.org/10.1126/sciadv.1701696.

(9) Seyler, K. L.; Rivera, P.; Yu, H.; Wilson, N. P.; Ray, E. L.; Mandrus, D. G.; Yan, J.; Yao, W.; Xu, X. Signatures of Moiré-Trapped Valley Excitons in MoSe2/WSe2 Heterobilayers. *Nature* **2019**, *567* (7746), 66–70. https://doi.org/10.1038/s41586-019-0957-1.

(10) Tran, K.; Moody, G.; Wu, F.; Lu, X.; Choi, J.; Kim, K.; Rai, A.; Sanchez, D. A.; Quan, J.; Singh, A.; Embley, J.; Zepeda, A.; Campbell, M.; Autry, T.; Taniguchi, T.; Watanabe, K.; Lu, N.; Banerjee, S. K.; Silverman, K. L.; Kim, S.; Tutuc, E.; Yang, L.; MacDonald, A. H.; Li, X. Evidence for Moiré Excitons in van Der Waals Heterostructures. *Nature* **2019**, *567* (7746), 71–75. https://doi.org/10.1038/s41586-019-0975-z.

(11) Jin, C.; Regan, E. C.; Yan, A.; Iqbal Bakti Utama, M.; Wang, D.; Zhao, S.; Qin, Y.; Yang, S.; Zheng, Z.; Shi, S.; Watanabe, K.; Taniguchi, T.; Tongay, S.; Zettl, A.; Wang, F. Observation of Moiré Excitons in WSe2/WS2 Heterostructure Superlattices. *Nature* **2019**, *567* (7746), 76–80. https://doi.org/10.1038/s41586-019-0976-y.

(12) Luo, Y.; Engelke, R.; Mattheakis, M.; Tamagnone, M.; Carr, S.; Watanabe, K.; Taniguchi, T.; Kaxiras, E.; Kim, P.; Wilson, W. L. In Situ Nanoscale Imaging of Moiré Superlattices in Twisted van Der Waals Heterostructures. *Nat. Commun.* **2020**, *11* (1), 4209. https://doi.org/10.1038/s41467-020-18109-0.

(13) Chen, W.; Sun, Z.; Wang, Z.; Gu, L.; Xu, X.; Wu, S.; Gao, C. Direct Observation of van Der Waals Stacking–Dependent Interlayer Magnetism. *Science (80-. ).* **2019**, *366* (6468), 983–987. https://doi.org/10.1126/science.aav1937.

(14) Hu, G.; Krasnok, A.; Mazor, Y.; Qiu, C.; Alù, A. Moiré Hyperbolic Metasurfaces. *Nano Lett.* **2020**, *20* (5), 3217–3224. https://doi.org/10.1021/acs.nanolett.9b05319.

(15) Baranov, D. G.; Xiao, Y.; Nechepurenko, I. A.; Krasnok, A.; Alù, A.; Kats, M. A. Nanophotonic Engineering of Far-Field Thermal Emitters. *Nat. Mater.* **2019**, *18* (9), 920–930. https://doi.org/10.1038/s41563-019-0363-y.

(16) Kats, M. A.; Sharma, D.; Lin, J.; Genevet, P.; Blanchard, R.; Yang, Z.; Qazilbash, M. M.; Basov, D. N.; Ramanathan, S.; Capasso, F. Ultra-Thin Perfect Absorber Employing a Tunable Phase Change Material. *Appl. Phys. Lett.* **2012**, *101* (22), 221101. https://doi.org/10.1063/1.4767646.

(17) Streyer, W.; Law, S.; Rooney, G.; Jacobs, T.; Wasserman, D. Strong Absorption and Selective Emission from Engineered Metals with Dielectric Coatings. *Opt. Express* **2013**, *21* (7), 9113. https://doi.org/10.1364/OE.21.009113.

(18) Cornelius, C. M.; Dowling, J. P. Modification of Planck Blackbody Radiation by Photonic Band-Gap Structures. *Phys. Rev. A* **1999**, *59* (6), 4736–4746. https://doi.org/10.1103/PhysRevA.59.4736.

(19) Lin, S.-Y.; Fleming, J. G.; Chow, E.; Bur, J.; Choi, K. K.; Goldberg, A. Enhancement and Suppression of Thermal Emission by a Three-Dimensional Photonic Crystal. *Phys. Rev. B* **2000**, *62* (4), R2243–R2246. https://doi.org/10.1103/PhysRevB.62.R2243.

(20) Narayanaswamy, A.; Chen, G. Thermal Emission Control with One-Dimensional Metallodielectric Photonic Crystals. *Phys. Rev. B* **2004**, *70* (12), 125101. https://doi.org/10.1103/PhysRevB.70.125101.

(21) Liu, X.; Tyler, T.; Starr, T.; Starr, A. F.; Jokerst, N. M.; Padilla, W. J. Taming the Blackbody with Infrared Metamaterials as Selective Thermal Emitters. *Phys. Rev. Lett.* **2011**, *107* (4), 045901. https://doi.org/10.1103/PhysRevLett.107.045901.

(22) Mason, J. A.; Smith, S.; Wasserman, D. Strong Absorption and Selective Thermal Emission from a Midinfrared Metamaterial. *Appl. Phys. Lett.* **2011**, *98* (24), 241105. https://doi.org/10.1063/1.3600779.

(23) Ikeda, K.; Miyazaki, H. T.; Kasaya, T.; Yamamoto, K.; Inoue, Y.; Fujimura, K.; Kanakugi, T.; Okada, M.; Hatade, K.; Kitagawa, S. Controlled Thermal Emission of Polarized Infrared Waves from Arrayed Plasmon Nanocavities. *Appl. Phys. Lett.* **2008**, *92* (2), 021117. https://doi.org/10.1063/1.2834903.

(24) Asano, T.; Suemitsu, M.; Hashimoto, K.; De Zoysa, M.; Shibahara, T.; Tsutsumi, T.; Noda, S. Near-Infrared–to–Visible Highly Selective Thermal Emitters Based on an Intrinsic Semiconductor. *Sci. Adv.* **2016**, *2* (12). https://doi.org/10.1126/sciadv.1600499.

(25) Greffet, J.-J.; Carminati, R.; Joulain, K.; Mulet, J.-P.; Mainguy, S.; Chen, Y. Coherent Emission of Light by Thermal Sources. *Nature* **2002**, *416* (6876), 61–64. https://doi.org/10.1038/416061a.

(26) Laroche, M.; Arnold, C.; Marquier, F.; Carminati, R.; Greffet, J.-J.; Collin, S.; Bardou, N.; Pelouard, J.-L. Highly Directional Radiation Generated by a Tungsten Thermal Source. *Opt. Lett.* **2005**, *30* (19), 2623. https://doi.org/10.1364/OL.30.002623.

(27) Dahan, N.; Niv, A.; Biener, G.; Gorodetski, Y.; Kleiner, V.; Hasman, E. Enhanced Coherency of Thermal Emission: Beyond the Limitation Imposed by Delocalized Surface Waves. *Phys. Rev. B* **2007**, *76* (4), 045427. https://doi.org/10.1103/PhysRevB.76.045427.

(28) Ma, W.; Alonso-González, P.; Li, S.; Nikitin, A. Y.; Yuan, J.; Martín-Sánchez, J.; Taboada-Gutiérrez, J.; Amenabar, I.; Li, P.; Vélez, S.; Tollan, C.; Dai, Z.; Zhang, Y.; Sriram, S.; Kalantar-Zadeh, K.; Lee, S.-T.; Hillenbrand, R.; Bao, Q. In-Plane Anisotropic and Ultra-Low-Loss Polaritons in a Natural van Der Waals Crystal. *Nature* **2018**, *562* (7728), 557–562. https://doi.org/10.1038/s41586-018-0618-9.

(29) Krasnok, A.; Baranov, D.; Li, H.; Miri, M.-A.; Monticone, F.; Alú, A. Anomalies in Light Scattering. *Adv. Opt. Photonics* **2019**, *11* (4), 892. https://doi.org/10.1364/aop.11.000892.

(30) Kinsey, N.; DeVault, C.; Boltasseva, A.; Shalaev, V. M. Near-Zero-Index Materials for Photonics. *Nat. Rev. Mater.* **2019**, *4* (12), 742–760. https://doi.org/10.1038/s41578-019-0133-0.

(31) Monticone, F.; Doeleman, H. M.; Hollander, W. Den; Koenderink, A. F.; Al, A.; Den Hollander, W.; Koenderink, A. F.; Alù, A. Trapping Light in Plain Sight: Embedded Photonic Eigenstates in Zero-Index Metamaterials. *Laser Photon. Rev.* **2018**, *12* (5), 1700220. https://doi.org/10.1002/lpor.201700220.

(32) Krasnok, A.; Alú, A. Embedded Scattering Eigenstates Using Resonant Metasurfaces. *J. Opt.* **2018**, *20* (6), 064002. https://doi.org/10.1088/2040-8986/aac1d6.

(33) Sakotic, Z.; Krasnok, A.; Cselyuszka, N.; Jankovic, N.; Alú, A. Berreman Embedded Eigenstates for Narrow-Band Absorption and Thermal Emission. *Phys. Rev. Appl.* **2020**, *13* (6), 064073. https://doi.org/10.1103/PhysRevApplied.13.064073.

(34) Sakotic, Z.; Stankovic, P.; Bengin, V.; Krasnok, A.; Alú, A.; Jankovic, N. Non-Hermitian Control of Topological Scattering Singularities Emerging from Bound States in the Continuum. *Laser Photon. Rev.* **2023**, 2200308. https://doi.org/10.1002/lpor.202200308.

(35) Zheng, Z.; Sun, F.; Xu, N.; Huang, W.; Chen, X.; Ke, Y.; Zhan, R.; Chen, H.; Deng, S. Tunable Hyperbolic Phonon Polaritons in a Suspended van Der Waals α-MoO3 with Gradient Gaps. *Adv. Opt. Mater.* **2022**, *10* (5), 1–9. https://doi.org/10.1002/adom.202102057.

(36) Passler, N. C.; Paarmann, A. Generalized 4 × 4 Matrix Formalism for Light Propagation in Anisotropic Stratified Media: Study of Surface Phonon Polaritons in Polar Dielectric Heterostructures. *J. Opt. Soc. Am. B* **2017**, *34* (10), 2128.

https://doi.org/10.1364/JOSAB.34.002128.

(37) Passler, N. C.; Jeannin, M.; Paarmann, A. Layer-Resolved Absorption of Light in Arbitrarily Anisotropic Heterostructures. *Phys. Rev. B* **2020**, *101* (16), 165425. https://doi.org/10.1103/PhysRevB.101.165425.

(38) Zanotto, S.; Mezzapesa, F. P.; Bianco, F.; Biasiol, G.; Baldacci, L.; Vitiello, M. S.; Sorba, L.; Colombelli, R.; Tredicucci, A. Perfect Energy-Feeding into Strongly Coupled Systems and Interferometric Control of Polariton Absorption. *Nat. Phys.* **2014**, *10* (11), 830–834. https://doi.org/10.1038/nphys3106.

(39) Baranov, D. G.; Wersäll, M.; Cuadra, J.; Antosiewicz, T. J.; Shegai, T. Novel Nanostructures and Materials for Strong Light-Matter Interactions. *ACS Photonics* **2018**, *5* (1), 24–42. https://doi.org/10.1021/acsphotonics.7b00674.

(40) Huang, Y.; Liu, Y.; Shao, Y.; Han, G.; Zhang, J.; Hao, Y. Rabi Splitting Obtained in a Monolayer BP-Plasmonic Heterostructure at Room Temperature. *Opt. Mater. Express* **2020**, *10* (9), 2159. https://doi.org/10.1364/ome.402194.

(41) Lepeshov, S.; Vyshnevyy, A.; Krasnok, A. Cloaking a Nanolaser. *arXiv:2110.14077 [physics.optics]* **2021**, 1–13.

(42) Sakotic, Z.; Krasnok, A.; Alu, A.; Jankovic, N. Topological Scattering Singularities and Embedded Eigenstates for Polarization Control and Sensing Applications. *Photonics Res.* **2021**, *9* (7), 1310–1323. https://doi.org/10.1364/prj.424247.